\documentclass[twocolumn,showpacs,preprintnumbers,amsmath,amssymb]{revtex4}

\usepackage{graphicx}
\usepackage{dcolumn}
\usepackage{bm}
\usepackage{natbib}

\def\yw{Y_{\rm w}}  \def\ew{\eta_{\rm w}}  \def\de{\Delta\eta} \def\bs{B_s}
\def\la{\langle} \def\ra{\rangle}
\def\bs{$\backslash$} 
 
\def\bd{\begin{document}} \def\ed{\end{document}}
\def\bmp{\begin{minipage}} \def\emp{\end{minipage}}
\def\bcc{\begin{center}} \def\ecc{\end{center}}     \def\npg{\newpage}
\def\beq{\begin{equation}} \def\eeq{\end{equation}} \def\hph{\hphantom}
\def\be{\begin{equation}} \def\ee{\end{equation}} \def\r#1{$^{[#1]}$}
\def\n{\noindent} \def\ni{\noindent} \def\pa{\parindent}
\def\hs{\hskip} \def\vs{\vskip} \def\hf{\hfill} \def\ej{\vfill\eject}
\def\cl{\centerline} \def\ob{\obeylines}  \def\ls{\leftskip}
\def\underbar#1{$\setbox0=\hbox{#1} \dp0=1.5pt \mathsurround=0pt
   \underline{\box0}$}   \def\ub{\underbar}    \def\ul{\underline}
\def\f{\left} \def\g{\right} \def\e{{\rm e}} \def\o{\over} \def\d{{\rm d}}
\def\vf{\varphi} \def\pl{\partial} \def\cov{{\rm cov}} \def\ch{{\rm ch}}
\def\la{\langle} \def\ra{\rangle} \def\EE{e$^+$e$^-$} \def\pt{p_{\rm T}}
\def\dt{\delta}
\def\bitz{\begin{itemize}} \def\eitz{\end{itemize}}
\def\btbl{\begin{tabular}} \def\etbl{\end{tabular}}
\def\btbb{\begin{tabbing}} \def\etbb{\end{tabbing}}
\def\beqar{\begin{eqnarray}} \def\eeqar{\end{eqnarray}}
\def\\{\hfill\break} \def\dit{\item{-}} \def\i{\item}
\def\bbb{\begin{thebibliography}{9} \itemsep=-1mm}
\def\ebb{\end{thebibliography}} \def\bb{\bibitem}
\def\bpic{\begin{picture}(260,240)} \def\epic{\end{picture}}
\def\akgt{\noindent{Acknowledgements}}
\def\fgn{\noindent{\bf\large\bf figure captions}}
\def\lan{\langle}
\def\ran{\rangle}
\def\p{\pi}
\def\ifmath#1{\relax\ifmmode #1\else $#1$\fi}%
\def\rc{\ifmath{{\mathrm{c}}}}
\def\cut{\ifmath{{\mathrm{cut}}}}
\def\rF{\ifmath{{\mathrm{F}}}}
\def\rK{\ifmath{{\mathrm{K}}}}
\def\rp{\ifmath{{\mathrm{p}}}}
\def\rt{\ifmath{{\mathrm{t}}}}
\def\LAB{\ifmath{{\mathrm{LAB}}}}
\def\cut{\ifmath{{\mathrm{cut}}}}
\def\beq{\begin{equation}}
\def\eeq{\end{equation}}
\def\us{^{(s)}}  \def\bea{\begin{eqnarray}} \def\eea{\end{eqnarray}}
\def\nbr{\nonumber} \def\e{\eta} \def\D{\Delta}
\def\r{\rho}  \def\unln{\underline}
\newcommand{\cinst}[2]{$^{\mathrm{#1}}$~#2\par}
\newcommand{\crefi}[1]{$^{\mathrm{#1}}$}
\newcommand{\crefii}[2]{$^{\mathrm{#1,#2}}$}
\newcommand{\crefiii}[3]{$^{\mathrm{#1,#2,#3}}$}
\newcommand{\HRule}{\rule{0.5\linewidth}{0.5mm}}
\usepackage{color}
\newcommand{\Blue}[1]{\textcolor[named]{Blue}{#1}}
\newcommand{\blue}[1]{\textcolor[named]{Blue}{#1}}
\newcommand{\Red}[1]{\textcolor[named]{Red}{#1}}
\newcommand{\red}[1]{\textcolor[named]{Red}{#1}}
\newcommand{\violet}[1]{\textcolor[named]{Violet}{#1}}
\newcommand{\brown}[1]{\textcolor[named]{Brown}{#1}}
\newcommand{\green}[1]{\textcolor[named]{Green}{#1}}
\newcommand{\magenta}[1]{\textcolor[named]{Magenta}{#1}}

\def\yw{Y_{\rm w}}  \def\ew{\eta_{\rm w}}  \def\de{\delta\eta} \def\bs{B_s} \def\ssnn{\sqrt {s_{NN}}}
\begin{document}


\title{Longitudinal boost-invariance of charge balance function\\
       in hadron-hadron and nucleus-nucleus collisions}

\author{Li Na, Li Zhiming, and Wu
Yuanfang} \affiliation{Key Laboratory of Quark and Lepton Physics
(Huazhong Normal University),Ministry of Education\\ Institute of
Particle Physics, Huazhong Normal University, Wuhan 430079, China}

\begin{abstract}
Using Monte Carlo generators of the PYTHIA model for hadron-hadron
collisions and a multi-phase transport (AMPT) model for
nucleus-nucleus collisions, the longitudinal boost-invariance of
charge balance function and its transverse momentum dependence are
carefully studied. It shows that the charge balance function is
boost-invariant in both {\it p}+{\it p} and Au+Au collisions in
these two models, consistent with experimental data. The balance
function properly scaled by the width of the pseudorapidity window
is independent of the position or the size of the window and is
corresponding to the balance function of the whole pseudorapidity
range. This longitudinal property of balance function also holds
for particles in small transverse momentum ranges in the PYTHIA
and the AMPT default models, but is violated in the AMPT with
string melting. The physical origin of the results are discussed.
\end{abstract}

\maketitle

\section{Introduction}

Charge balance function (BF) has been widely used as an effective
exploring for the hadronization scheme in hadron-hadron collisions
at the ISR energies~\cite{oldbf1} and $e^{+} + e^{-}$
annihilations at PETRA energies~\cite{petra}. Recently, the charge
BF gains special attentions in clocking hadronization at
relativistic heavy-ion collisions. A narrowing of the BF is
suggested as a signature for delayed hadronization~\cite{bf1,bf2}.

The dependence of the BF on centrality and system size has been
reported by several relativistic heavy-ion
experiments~\cite{star130,na49}. However, most of the current
heavy-ion experiments are limited by the pseudorapidity
range~\cite{star130,na49,phenix}, it is impossible to
quantitatively compare the results from the experiments with the
coverage at different pseudorapidity ranges. The dependence of the
BF on the pseudorapidity window is essential for understanding the
physics of the BF~\cite{star130,na49,na22bf,tom1}, and has been
carefully studied by NA22~\cite{na22bf} and STAR experiments for
hadron-hadron and relativistic heavy ion collisions, respectively.

The NA22 experiment has full 4$\pi$ acceptance and excellent
momentum resolution~\cite{na22bf}. It is found in the experiment
that the BF in $\pi^{+}\rp$ and $\rK^{+}\rp$ Collisions at 22 GeV
is invariant under longitudinal boost over the whole rapidity
range of produced particles, in spite of the non-boost-invariance
of the single-particle density. Moreover, the BF of the whole
rapidity range can be deduced from the BF properly scaled by the
width of rapidity windows~\cite{na22bf}.

The STAR experiment covers a finite but relative wide
pseudorapidity range. The scaling property of the BF in Au +Au
collisions at 200 GeV is further observed in the
experiment~\cite{star200}. This scaling property of the balance
function is also found in different $\pt$ ranges of final state
particles.

These results from both hadron-hadron and nuclear collisions
indicate that charge balance of produced particles in strong
interactions is boost-invariance in longitudinal phase-space, in
contrary with the single particle density.

Therefore, it is interesting to see if those properties are taken
into account in the models which are successfully described
hadron-hadron and nuclear collisions, and how they associate with
the mechanisms of particle production in the models.

\section{Charge balance function and implement models}

Charge balance function measures how the conserved electric charge
compensate in the phase space, i.e., how the surrounding net
charge are rearranged if the charge of a selected point
changes\cite{oldbf1}. In high energy collisions, the production of
charged particles are constrained by charge balance in the phase
space. The BF therefore provides a direct access to collision
dynamics.

The BF has been originally defined in terms of a combination of
four kinds of charge-related conditional densities in
pseudorapidity~\cite{oldbf1}
\begin{eqnarray}
B(\eta_1|\eta_2)&=&
\frac{1}{2}\left[\r(+,\eta_1|-,\eta_2)-\r(+,\eta_1|+,\eta_2)\right.\nonumber \\
&& \quad \left.+\r(-,\eta_1|+,\eta_2)-\r(-,\eta_1|-,\eta_2)\right]
,
\end{eqnarray}
where the notation $\r(a,\eta_a|b,\eta_b)$ represents the ratio $
\r_{ab}(\eta_a,\eta_b) / \r_b(\eta_b) = \lan n_{ab}(\eta_a,
\eta_b) \ran / \lan n_b(\eta_b) \ran $ with $a, b$ standing for
$+$ or $-$ charged particles. Projecting to pseudorapidity
difference $\de =\eta_1-\eta_2$ in an pseudorapidity window
$\eta_w$, it becomes~\cite{bf1,star130}
\begin{eqnarray}
 B(\de|\ew) &=& \frac{1}{2}\left[
 \frac{\la n_{+-}(\de)\ra-\la n_{++}(\de)\ra}{\la
 n_+\ra}\right.
 \nonumber \\
 & &\quad \left.+\frac{\la n_{-+}(\de)\ra-\la n_{--}(\de)\ra}{\la
n_-\ra}\right]
\end{eqnarray}
where $n_{ab}(\de)$ is the total number of pairs of opposite
charged particles with pseudorapidity difference $\de$ in the
pseudorapidity window $\ew$. $n_+$ and $n_-$ are the number of
positively and negatively charged particles in the window $\ew$,
respectively. $\lan\cdots\ran$ is the average over the whole event
sample.

From the findings of the BF at NA22~\cite{na22bf} and STAR
experiments~\cite{star200}, the BF is boost-invariant in the whole
rapidity range in hadron-hadron collisions and may be in nuclear
collisions as well. In the case, the properly scaled BF is
corresponding to the BF of the whole pseudorapidity range and is
deduced by
\begin{eqnarray}
B_s(\delta\eta)=\frac{B(\delta\eta|\eta_{\rm
w})}{1-\frac{\delta\eta}{|\eta_{\rm w}|}}
\end{eqnarray} \noindent
where $|\eta_{\rm w}|$ is the width of pseudorapidity window.

The PYTHIA 5.720~\cite{pythia} is well set up for {\it p}+{\it p}
collisions. It is a standard Monte Carlo generator with string
fragmentation as hadronization scheme. Two versions of a
multi-phase transport (AMPT) model~\cite{ampt} are used to study
Au+Au collisions. One is the AMPT default and the other one is the
AMPT with string melting. In both versions, the initial conditions
are obtained from the HIJING model, and then the scattering among
partons are given by ZPC. In the AMPT default model, the partons
recombined with their parent strings when they stop interacting,
and the resulting strings are converted to hadrons using the Lund
string fragmentation model, while in the AMPT model with string
melting, quark coalescence is used in combining partons into
hadrons. The dynamics of the hadronic matter is described by ART
model.

It is commonly believed that in relativistic heavy ion collisions,
the charge ordering during the string fragmentation in elementary
collisions is no longer valid, and it should be replaced by the
quark-coalescence mechanism in hadronization
~\cite{recombination}. So it is interesting to see whether the
boost-invariance of the BF is sensitive to the mechanisms of
hadronization.

In this paper, we firstly study the boost-invariance of the BF for
{\it p}+{\it p} collisions at $\sqrt s= 22$ GeV and $\sqrt s= 200$
GeV using the PYTHIA, and for Au+Au collisions at $\sqrt s= 200$
GeV using two versions of the AMPT. The transverse momentum
dependence of longitudinal scaling property of the BF is then
examined in the models. The obtained results are compared with
corresponding experimental data and discussed.

\section{Boost-invariance and longitudinal scaling of the BF}
\begin{figure}
\includegraphics[scale=0.45]{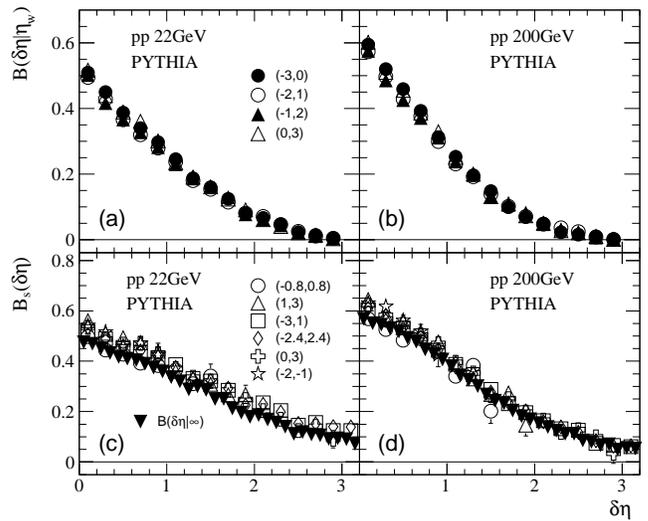}
\caption{\label{fig:epsart} Upper panel: the
$B(\delta\eta|\eta_{\rm w})$ in four pseudorapidity windows with
equal size $|\eta_{\rm w}|=3$ at the different positions for {\it
p}+{\it p} collisions at (a) $\sqrt{s}=22$ GeV and (b)
$\sqrt{s}=200$ GeV by PYTHIA model. Lower panel: the scaled
balance function, $B_s(\delta\eta)$, deduced from the directly
measured BF at six different sizes and positions of pseudorapidity
windows for {\it p}+{\it p} collisions at (c) $\sqrt{s}=22$ GeV
and (d) $\sqrt{s}=200$ GeV by PYTHIA model. The solid down
triangle is the BF of the whole $\eta$ range.}
\end{figure}

In order to demonstrate directly whether the BF is invariant under a
longitudinal Lorentz transformation over the whole rapidity in
hadron-hadron collisions, we choose four equal size ($|\eta_{\rm
w}|=3$) pseudorapidity windows locating at different positions
($-3,0$), ($-2,1$), ($-1,2$) and ($0,3$). The results for {\it
p}+{\it p} collisions at $\sqrt s= 22$ GeV and $\sqrt s= 200$ GeV
are shown in Fig.~1(a) and (b) respectively. The statistic errors
are smaller than the markers. It is clear that the BF measured in
four windows are approximately identical to each other at two
incident energies. This indicates that the charge compensation is
essentially the same in any longitudinally-Lorentz-transformed frame
for {\it p}+{\it p} collisions in the PYTHIA model, consistent with
the data from NA22 experiment. These results show that the string
fragmentation mechanism implemented in PYTHIA well describes the
production mechanisms of charge particles and their charge balance
in longitudinal phase space.

Fig.~1(c) and (d) are the scaled balance function
$B_s(\delta\eta)$ at two incident energies. They are deduced from
directly measured $B(\delta\eta|\eta_{\rm w})$ at six different
pseudorapidity windows, ($-0.8,0.8$) (open circles), ($1,3$) (open
triangles), ($-3,1$) (open squares), ($-2.4,2.4$) (open diamonds),
($0,3$) (open crosses), and ($-2,-1$) (open stars). From the
figures we can see that all the $B_s(\delta\eta)$ deduced from
different windows are coincide with each other within errors, as
expected from boost-invariance of the BF~\cite{bf2}. The solid
down triangles in the same figures are the BF of the whole
pseudorapidity range, $B(\delta\eta|\eta_\infty)$. It is close to
the scaled balance function $B_s(\delta\eta)$. These results
indicate that the scaled BF is in fact corresponding to the BF of
the whole pseudorapidity range $B(\delta\eta|\infty)$~\cite{bf2}.

\begin{figure}
\includegraphics[scale=0.45]{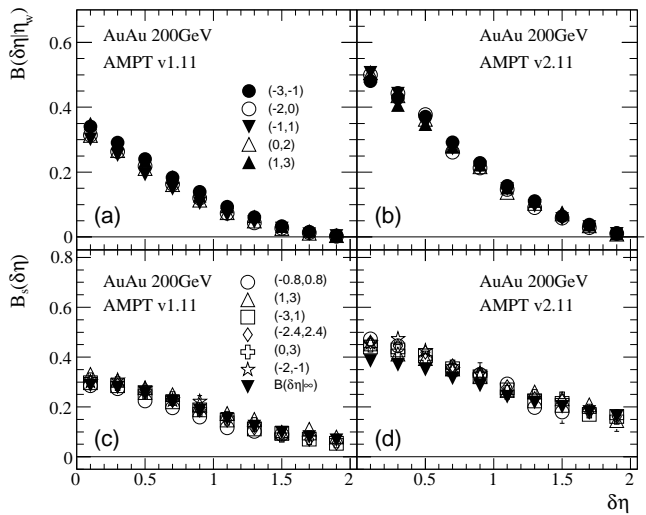}
\caption{\label{fig:epsart} Upper panel: the
$B(\delta\eta|\eta_{\rm w})$ in five pseudorapidity windows with
equal size $|\eta_{\rm w}|=2$ at the different positions for Au+Au
collisions at $\sqrt{s}=200$ GeV by (a) the AMPT default and (b)
the AMPT with string melting. Lower panel: the scaled balance
function, $B_s(\delta\eta)$, deduced from the directly measured BF
at various pseudorapidity windows with different sizes and
positions for Au+Au collisions at $\sqrt{s}=200$ GeV by (c) the
AMPT default and (d) the AMPT with sting melting. }
\end{figure}

It is then interesting to see whether the boost-invariance of the
BF is held in nucleus-nucleus collisions. STAR experiment only
observe the boost-invariance of BF in cental pseudorapidity range
$-1<\eta<1$~\cite{star200}, where the single particle distribution
is almost flat, or boost-invariance. Now in model investigation,
we can carefully examine the property in the whole pseudorapidity
range.

The upper panel of Fig.~2 is the BF in five pseudorapidity windows
with equal size $\eta_{\rm w}=2$ at different positions ($-3,-1$),
($-2,0$), ($-1,1$), ($0,2$) and ($1,3$). Where the Fig.~2(a) and
(b) are the results from the AMPT default (v1.11) and the AMPT
with string melting (v2.11), respectively. Both figures show that
the BF is boost-invariance in pseudorapidity range (-3, 3) in two
versions of the AMPT.

The lower panel of Fig.~2 is the scaled balance functions, which are
obtained from directly measured BF at six different windows as
indicated at legend of the figure, where the solid down triangles
are the BF in pseudorapidity range (-4, 4). It shows that the scaled
BF does not depend on the size and position of the windows, and
corresponds to the BF of the whole pseudorapidity in two versions of
the AMPT, consistent with the results of {\it p}+{\it p} collisions
in the PYTHIA model.

\section{The transverse-momentum dependence of the boost-invariance of the BF}

The longitudinal property of boost invariance of BF comes from the
special longitudinal interaction of charged particles under the
constraint of global electric charge balance. Global electric
charge conservation not only applies to all final-state charged
particles, but also constrains particles which are produced at the
same proper time of evolution. It is argued that the
transverse-momentum of final-state particles may be roughly used
as a scale of the proper time of their production in the expansion
of nuclear collisions~\cite{rudy, bmuller, hama1, ptscale}.
Examining the $p_{\rm T}$ dependence of longitudinal property of
the BF will provide direct access on whether particles in
specified $p_{\rm T}$ range are consistent to be produced
simultaneously with well balanced electric charge.

So we turn to check whether the boost-invariant of BF holds for
particles in different $\pt$ ranges. Fig.~3 shows the BF for {\it
p}+{\it p} collisions at $\sqrt s= 22$ GeV and $\sqrt s= 200$ GeV
from PYTHIA in three transverse momentum bins ($0<\pt<0.2$),
($0.2<\pt<0.4$), and ($\pt>0.2$) GeV/$c$, respectively. These $\pt$
bins are selected to make the multiplicity in each bin comparable.
The result shows that the points at a given $\delta\eta$ in a
restricted $p_{\rm T}$ interval are approximately coincide with each
other, i.e., the boost-invariance of the BF hold in small $p_{\rm
T}$ ranges. It indicates that particles produced at different $\pt$
ranges are also boost-invariant for hadron-hadron collisions in the
PYTHIA model.

\begin{figure*}
\includegraphics[width=12cm]{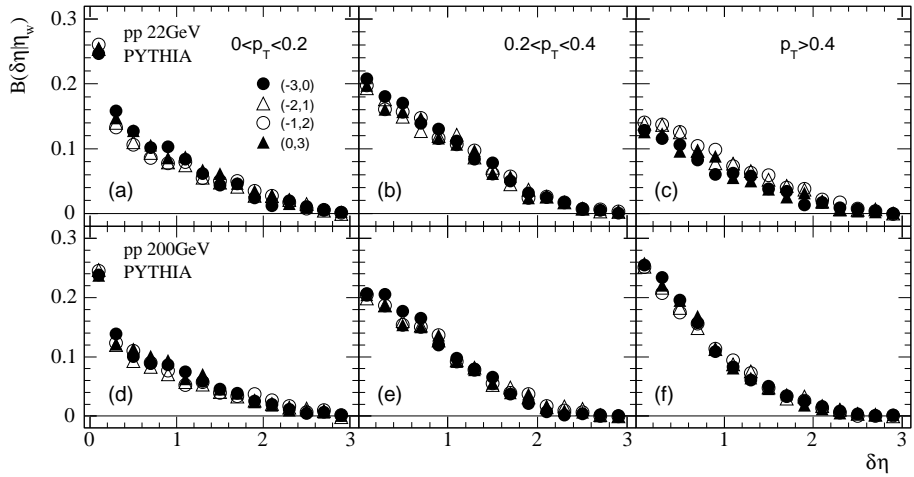}
\caption{\label{fig:epsart}For each of three $p_{\rm T}$ ranges,
the $B(\delta\eta|\eta_{\rm w})$ in four pseudorapidity windows
with equal size $|\eta_{\rm w}|=3$ at the different positions for
{\it p}+{\it p} collisions at $\sqrt{s}=22$ GeV and $\sqrt{s}=200$
GeV in upper and lower panels, respectively.}
\end{figure*}

\begin{figure*}
\includegraphics[width=16cm]{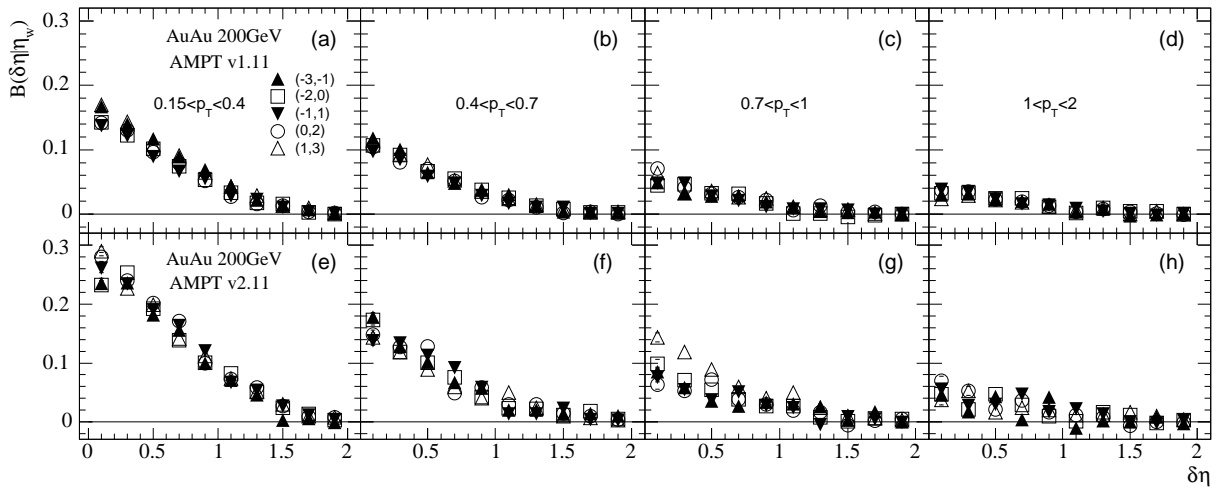}
\caption{\label{fig:epsart}  For each of four $\pt$ ranges, the
$B(\delta\eta|\eta_{\rm w})$ in five pseudorapidity windows with
equal size $|\eta_{\rm w}|=2$ at the different positions for Au+Au
collisions at $\sqrt{s}=200$ GeV from the AMPT default (in upper
panel) and the AMPT with string melting (in lower panel).}
\end{figure*}

The same study for Au+Au 200 GeV collisions from the two versions of
the AMPT are presented in the upper and lower panels of Fig.~4,
respectively. Where four $\pt$ bins are, ($0.15, 0.4$), ($0.4,
0.7$), ($0.7, 1$) and ($1, 2$) GeV/$c$. From the upper panel of the
figure, we can see that the BF of different pseudorapidity windows
in each $\pt$ bin are close to each other, in consistent with the
data from STAR experiment~\cite{star200}. However, in the AMPT with
string melting, as shown in the lower panel of the figure, where the
BF of different pseudorapidity windows are not as close to each
other as those in the upper panel.

This is because in the AMPT with string melting, each parton in
the evolution of nuclear collision has its own freeze-out time,
which last a very long period after the interaction of two
nucleus~\cite{liu-yu}. The particles in the same
transverse-momentum range are not freezed-out simultaneously with
well balanced charge, and therefore the longitudinal
boost-invariance of the BF in small $\pt$ ranges is violated. In
the AMPT default, the partons recombined with their parent strings
immediately after they stop interacting, and converted to hadrons.
So the charge balance of the produced particles in the same $\pt$
ranges is preserved and boost-invariance of the BF keeps.

\section{Summary}

In the paper, we systematically study the longitudinal
boost-invariance of charge balance function and its $p_{\rm T}$
dependence for {\it p}+{\it p} and Au+Au collisions using PYTHIA the
AMPT models. It shows that charge balance function is
boost-invariance in both hadron-hadron and nuclear interactions, in
contrary to the single particle density. As expected, this
boost-invariance of the BF make the BF properly scaled by window
size is independent of window and corresponds to the BF of the whole
(pseudo)rapidity range. Therefore, the BF is a good measure free
from the restriction of finite longitudinal acceptance.

It is further show that the boost invariance of the BF in
specified $p_{\rm T}$ range is valid in PYTHIA for hadron-hadron
collisions and the AMPT default for Au+Au collisions. While the
AMPT with string melting fails to reproduce this property due to
the different schemes at hadronization. So the $\pt$ dependence of
the longitudinal property of the BF may be served as a sensitive
probe for charge balance in hadronization mechanism.

\section{Acknowledgments}
We thank Prof. Liu Lianshou and Dr. Yu
Meiling for valuable discussions and remarks. This work is
supported in part by the NSFC of China with project No. 10835005
and No. 10647124, and the MOE of China with project No. IRT0624
and No. B08033.


\end{document}
%